\documentclass[]{pasj01}

\begin{document} 
\Received{}
\Accepted{}

\title{The nature of the X-ray pulsar in M31: an intermediate mass X-ray binary?}

\author{Shigeyuki \textsc{Karino}\altaffilmark{1}}%
\altaffiltext{1}{Kyushu Sangyo Univ. Dept. of Engineering, Matsukadai 2-3-1, Fukuoka 813-8503, Japan}
\email{karino@ip.kyusan-u.ac.jp}




\maketitle

\begin{abstract}

Recently the first finding of a spin period of an accreting neutron star in M31 is reported. 
The observed spin period is 1.2 s and it shows 1.27 d modulations due to orbital motion. 
From the orbital information, the mass donor could not be a giant massive star. 
On the other hand, the observed properties are quite odd for typical low mass X-ray binaries.

In this study, we compare observed binary parameters with theoretical models given by a stellar evolution track and make a restriction on the possible mass range of the donor. 
According to the standard stellar evolution model, the donor star should be larger than 1.5 $M_{\odot}$, and this suggests that this system is a new member of a rare category, intermediate mass X-ray binary. 
The magnetic field strength of the neutron star suggested by spin-up/down tendency in this system supports the possibility of intermediate mass donor.

\end{abstract}

\section{Introduction}

Recently, in EXTraS project with XMM Newton \citep{D15}, periodic pulsations from an X-ray point source, 3XMM \, J004301.4+413017 (hereafter, we refer as gM31 XBPh), in an external arm of M31 have been found \citep{E15}. 
This object shows 1.2 second periodic pulsations with clear single peak, associated with 1.27 days orbital modulations. 
From these evidences, it is understood that this system is an X-ray binary system with an accreting neutron star. 
Though there have been many systematic surveys of X-ray sources in M31 \citep{SG09,S11}, this is the first finding of a spin period of a neutron star in accreting X-ray binary systems in this galaxy. 
The X-ray luminosity of this object is fluctuating: the maximum luminosity is $2 \times 10^{38} \rm{erg \, s}^{-1}$ and this luminosity is close to the Eddington luminosity of a neutron star \citep{ST83}.

Since this object shows a large luminosity and a hard X-ray emission, it has been supposed to be a member of high mass X-ray binary (HMXB) in M31 \citep{SG09}.
However, since the position of this object is just a neighbor of a globular cluster, also there is a possibility that this system is one of low mass X-ray binaries (LMXBs), which is popular also in M31 \citep{S11,Z16}. 
Currently we have at least some indications that the mass donor in this system is not a massive star.
For instance, the orbital period is shorter than most of HMXBs \citep[for example]{K07}.
Furthermore, since there are no significant occultation and/or dipping, a massive donor with a large radius is probably rejected \citep{E15}. 
Additionally, the optical observations, a bright optical counterpart with $M_{\rm{V}} < -2.5$ is restricted and this limit also disfavors a high mass donor \citep{E15}. 
Because of these observed evidences, the mass donor of M31 XBP would not be a massive star.
At this time, however, the actual donor type of this system has not been confirmed: the remained possibilities of this system are (1) a LMXB system in the globular cluster, as like the slowly rotating LMXB 4U 1626-67 \citep{JK01,Z16}, or (2) intermediate mass X-ray binary (IMXB) system as like a Her X-1 \citep{T72, B97} overlapped with a globular cluster by chance
\footnote{
In the past, since the sample of X-ray pulsars had been small, an X-ray binary system which donor mass is less than $2.5 M_{\odot}$ had been categorized as a LMXB \citep{B97}. 
Here, however, in accordance with \citet{E15}, we designate a system which donor mass is between $\approx 1 M_{\odot}$ to $\approx5 M_{\odot}$ as an IMXB.
}.

In this study, we discuss the true nature of M31 XBP by using a theoretical stellar evolution track and magnetic properties of a neutron star. 
Our goal is to limit the type of mass donor in this system, only from the observed X-ray data.
In the next section, we introduce our tactics. 
In Section 3, we show that the mass donor in this system cannot be a massive star, while too small star also cannot be allowed. 
In the discussion, we examine the magnetic properties of the neutron star in M31 XBP to support our result. 
We propose that a mass donor with intermediate mass up to $2 M_{\odot}$ could satisfy all the conditions. 
The final section is devoted to conclusions.

\section{Roche Lobe Over Flow Accretion}

In order to investigate the true nature of M31 XBP, first we focus on the radius of the mass donor in this binary system. 
The radius of the donor is directly concerned with the accretion mode in a close binary system. 
In general, LMXB is powered by mass accretion from the donor to the compact object (in this study, we consider only a neutron star) via Roche lobe overflow (RLOF). 
Also a capture process of strongly enhanced stellar wind due to X-ray irradiation can be considered as a power source of LMXB \citep{ITF97}. 
However, in order to enhance the stellar wind of a low mass donor, considerably small orbit ($P_{\rm{orb}} \sim $ a few hr) is required. 
Additionally, even in such a tight system, the small mass donor should fill more than 80\% of its Roche lobe in order to supply enough mass via stellar wind \citep{ITF97}.
Hence, even in this case a small donor in LMXB should substantially fill its Roche lobe. 
Hereafter, we consider only RLOF case as an engine of LMXB. 

If M31 XBP is really a LMXB and a RLOF accretion is setting on, the mass donor should have an enough large radius to fill its Roche lobe. 
The approximated Roche lobe radius is given by \citet{E83} by the following formula:
\begin{equation}
R_{\rm{RL}} = \frac{0.49 q^{2/3}}{0.6 q^{2/3} + \ln \left(1+ q^{1/3} \right) } a,
\label{eq:Rrl}
\end{equation}
where $q = M / M_{\rm{NS}}$ is the mass ratio between the donor star and the accreting neutron star. 
Here, $a$ is the orbital semi-major axis;
\begin{equation}
a = \left[ \frac{P_{\rm{orb}}^2}{4 \pi^2} G \left( M + M_{\rm{NS}} \right) \right] ^{1/3} .
\label{eq:a}
\end{equation}
From the observation, the orbital eccentricity of M31 XBP is indicated as almost zero \citep{E15}. 
Since the variation of the neutron star mass affects $a$ and $R_{\rm{RL}}$ only slightly, we use a fixed value, $M_{\rm{NS}} = 1.4 M_{\odot}$.

Stellar radius can be obtained from an appropriate stellar evolution track as a function of stellar age and initial mass.
In this study, we use an approximated stellar evolution track given by \citet{HPT00} (hereafter, HTP2000).
Here, we compute evolutions of stellar radii from the zero-age main sequence (ZAMS) phase given by \citet{T96}, to the beginning of giant branch (BGB) phase, by implementing equations (1) to (30) in HPT2000. 
A stellar evolution is primarily determined by its ZAMS mass; a disturbance from its companion is not so significant.
Hence, though the procedure in HTP2000 gives a single stellar evolution, 
it has been broadly used in binary evolution codes \citep{H02,B02}.
If the stellar radius achieves the Roche lobe radius before the terminal main sequence (TMS) phase, or subsequent Hertzsprung gap stage, the outer envelope of the star overflows from its Roche lobe and falls onto the compact object via the first Lagrangian point. 
In this case, the overflow proceeds within almost thermal time scale, 
\begin{eqnarray} 
\tau_{\rm{th}} &\sim& \frac{GM^2}{RL} \nonumber \\
&=& 3.13 \times 10^{7} \left( \frac{M}{M_{\odot}} \right)^{2}
\left( \frac{R}{R_{\odot}} \right)^{-1} \left( \frac{L}{L_{\odot}} \right)^{-1} \rm{yr}, 
\label{eq:tauth} 
\end{eqnarray} 
and the mass accretion goes stably \citep{KW96}. 
The mass transfer rate estimated from this time scale is
\begin{eqnarray}
\dot{M} &\sim& \frac{M}{\tau_{\rm{th}}} = \frac{RL}{GM}  \nonumber \\
&=& 2.01 \times 10^{18} \left( \frac{M}{M_{\odot}} \right)^{-1}
\left( \frac{R}{R_{\odot}} \right) \left( \frac{L}{L_{\odot}} \right) \rm{g \, s}^{-1}.
\end{eqnarray}
Assuming that the potential energy of this transferred mass is converted into X-rays at the neutron star surface, the corresponding luminosity becomes
\begin{eqnarray}
L_{\rm{X}} &\sim& \frac{G M_{\rm{NS}} \dot{M}}{R_{\rm{NS}}}  \nonumber \\
&=& 3.74 \times 10^{38} 
\left( \frac{M}{M_{\odot}} \right)^{-1}
\left( \frac{R}{R_{\odot}} \right) \left( \frac{L}{L_{\odot}} \right) \rm{erg \, s}^{-1}.
\label{eq:lx}
\end{eqnarray}
Actually, only a few 10\% of overflowed mass from the donor can be accreted onto the compact component \citep{W11}. 
Additionally, about 10\% of the potential energy of accreted matter can be converted into X-rays at the surface of the neutron star \citep{ST83}.
Therefore, we preferably need to add a factor of $\mathcal{O} (0.1)$ (this factor may range from 0.01 to 1) in this luminosity estimation.

\begin{figure}
 \begin{center}
  \includegraphics[width=8cm]{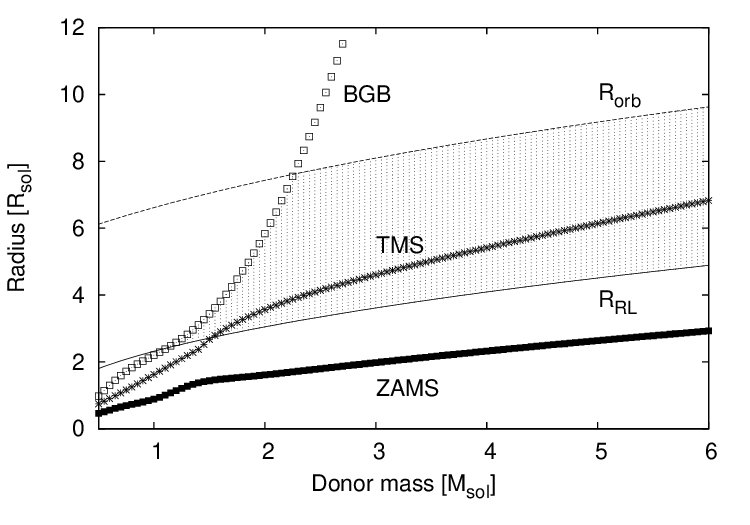} 
 \end{center}
\caption{
The RLOF condition (shown by shaded area). 
We show the orbital radius given by Eq.~(\ref{eq:a}) (dashed curve) and the Roche lobe radius given by Eq.~(\ref{eq:Rrl}) (solid curve).
At the same time, stellar radii of the mass donor as functions of stellar mass are shown at ZAMS phase (filled squares), TMS phase (asterisks), and BGB phase (open squares), respectively.
}\label{fig:1}
\end{figure}

If the stellar radius reaches its Roche lobe after evolving to a giant, on the other hand, the stellar envelope becomes highly convective and RLOF proceeds in the dynamical timescale. 
In this case, the binary system rather forms a common envelope. 
During the common envelope phase, a lot of stellar mass spreads around the binary, and X-rays from the compact companion would be hardly observed \citep{I13}. 
Then, the condition that RLOF works is, 
\begin{equation}
R_{\rm{RL}} < R < R_{\rm{BGB}},
\end{equation}
where $R_{\rm{BGB}}$ denotes the corresponding donor radius at the BGB phase. 
Of course, the stellar radius $R$ should be smaller than the orbital separation $a$, we have to impose a strong upper condition, 
\begin{equation}
R < a .
\end{equation}
We show these conditions in Fig.~\ref{fig:1}.
In this figure, we show the orbital radius given by Eq.~(\ref{eq:a}) (dashed curve) and the Roche lobe radius given by Eq.~(\ref{eq:Rrl}) (solid curve) for the $P_{\rm{orb}} = 1.27 \rm{d}$ system containing a $1.4 M_{\odot}$ neutron star. 
In this figure, at the same time, stellar radii of the mass donor as functions of stellar mass are shown for three selected evolutionary phases: ZAMS phase (filled squares), TMS phase (asterisks), and BGB phase (open squares). 
Comparing these radii, we can evaluate the possible mass range where RLOF can be set on. 
Then the shaded region between two curves indicates the possible condition of RLOF accretion.

\section{Result}

In the first report by \citet{E15}, the minimum mass of the donor in M31 XBP is estimated as 0.4 $M_{\odot}$. 
Here, taking stellar evolutions into account, we try to make further strong restrictions on the mass of the donor.
First of all, the lightest stars ($< 0.95 M_{\odot}$) are ruled out, because they cannot evolve and expand until Roche lobe filling radius within the cosmic age. 
For instance, for a small star with $M = 0.8 M_{\odot}$, it takes 27.9 Gyr to evolve to a red giant star, and it is much longer than the cosmic age. 
Furthermore, a donor star the mass of which is less than $1.1 M_{\odot}$ cannot fill its Roche lobe until it evolves to a red giant stage, in a $P_{\rm{orb}} = 1.27 \rm{d}$ binary system.
Even if it fills the Roche lobe, however, a convective outer envelope sets in when a sun-like star evolves to a red giant. 
With such a convective envelope, the RLOF will proceed within a dynamical time-scale, and in this case matter cannot be captured by a compact companion completely. 
Then, the system rather forms a common envelope and X-rays from a compact star will be obscured. 
Consequently, such smaller stars with $M < 1.1 M_{\odot}$ could be ruled out for an appropriate donor of M31 XBP. 
This is stronger condition than one suggested by \citet{E15} ($M > 0.4 M_{\odot}$). 

In \citet{E15}, it is suggested that M31 XBP could be an intermediate mass X-ray binary.
Her X-1 is one of the few examples of such intermediate mass X-ray binaries, with an estimated donor mass $\sim 2 M_{\odot}$. 
The neutron star spin period and orbital period of Her X-1 are $P_{\rm{spin}} = 1 \rm{s}$ and $P_{\rm{orb}} = 1.7 \rm{d}$, respectively \citep{T72}, and these data are similar to the present binary in M31.
Examining Fig.~\ref{fig:1}, stars with mass in the range between $1.1 M_{\odot}$ to $1.5 M_{\odot}$ can fill the Roche lobe in the Hertzsprung gap stage after the terminal main sequence. 
While on the other hand, stars with larger mass than $1.5 M_{\odot}$ can satisfy RLOF condition during the main sequence phase. 
Furthermore, stars heavier than $2.3 M_{\odot}$ would expand to the binary orbital radii during their Hertzsprung gap phases. 
Since, in general, the time-scale of the Hertzsprung gap is much shorter than the stellar life-time, the observational possibility would become lower for the case that the Roche radius is filled during the Hertzsprung gap phase. 
For example, a $M = 2.3 M_{\odot}$ star ends its main sequence evolution in 780Myr, and within this duration, RLOF condition is satisfied for 140Myr. 
On the other hand, its Hertzsprung gap stage ends within only 5.7Myr. 
It statistically suggests that the donor is still on the main sequence, if its mass is larger than $2 M_{\odot}$.

\begin{figure}
 \begin{center}
  \includegraphics[width=8cm]{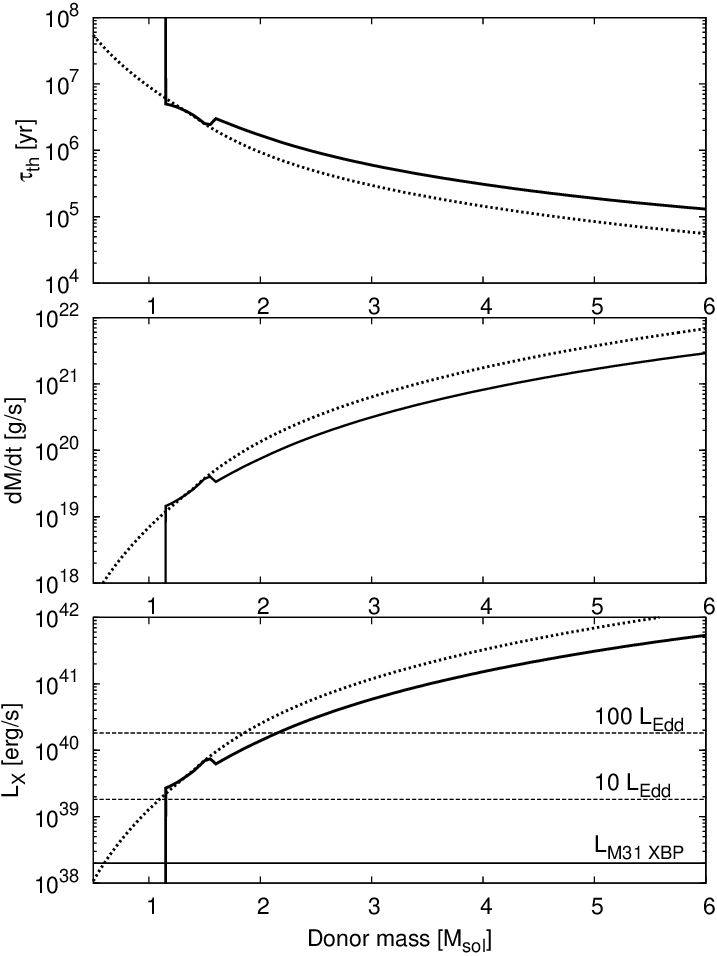} 
 \end{center}
\caption{
Accretion properties of RLOF.
We show the mass transfer time (thermal time) in upper panel, the mass transfer rate in the middle panel, and the maximum X-ray luminosity in the lower panel, respectively.
The solid curves in these figures show the parameters evaluated at the point where the donor radius reaches the Roche radius. 
On the other hand, the dotted curves show them evaluated at the TMS stage. 
In the lower panel, we also show the observed X-ray luminosity of M31 XBP, with 10 and 100 times of Eddington luminosity by horizontal lines.  
}\label{fig:2}
\end{figure}

Up to this point, the upper limit on the donor mass is yet to be determined. 
For a loose condition, as argued by \citet{E15}, the donor radius cannot be so large, 
since there is no occultation. 
The luminosity limit of the donor, $M_{\rm{V}} > -2.5$, also makes a rough restriction on the donor mass.
From the observed X-ray data, we could impose a further condition on the donor mass in this system.
Here we assume that a star which mass is a few $M_{\odot}$ fills its Roche lobe.
Then, the mass transfer time (thermal time), the mass transfer rate, and the maximum X-ray luminosity are given by Eqs.~(\ref{eq:tauth}) -- (\ref{eq:lx}).
In Fig.~\ref{fig:2}, we show these values obtained by substituting stellar parameters given by the stellar evolution track of HPT2000.
We show the thermal time in the upper panel, the mass transfer rate in the middle panel, and the maximum X-ray luminosity in the lower panel, respectively. 
The solid curves in these figures show the parameters evaluated at the point where the donor radius reaches the Roche radius. 
On the other hand, the dotted curves show them evaluated at the TMS stage. 
In the lower panel, we also show the observed X-ray luminosity of M31 XBP, with 10 and 100 times of Eddington luminosity by horizontal lines.  
From the upper panel, we can see that the thermal time becomes shorter for larger donor mass. 
The thermal time becomes shorter than $10^{6} \rm{yr}$ when the donor mass exceeds $2.4 M_{\odot}$. 
When donor mass achieves $5 M_{\odot}$, it becomes as short as $2 \times 10^{5} \rm{yr}$, and it falls below $10^{5} \rm{yr}$ when $M \geq 6.9 M_{\odot}$.
Such short time-scales of accretion duration would decrease the chance of observation. 
Furthermore, from the middle and lower panels, if the donor mass is larger than $2 M_{\odot}$, the mass transfer rate and the X-ray luminosity reaches 100 times larger than the Eddington rate. 
It means that the donor mass of $2 M_{\odot}$ would be consistent if the efficiencies of mass accretion and energy conversion give a factor 0.01 in Eq.~(\ref{eq:lx}).
In actuality, the mass transfer efficiency of RLOF would be a few 10\% \citep{W11}, and the emission efficiency of an accreting neutron star would be $\gtrsim 0.1$ \citep{ST83}. 
Considering a efficiency factor of 0.1 in Eq.~(\ref{eq:lx}), $M \approx 1.5 M_{\odot}$ would be most likely consistent with the observed X-ray luminosity, and possibly up to $2 M_{\odot}$. 
From these points of view, it is strongly suggested that the donor of this system is intermediate mass star around $\approx 1.5 M_{\odot}$, as like Her X-1.

We have to note that the estimation of the mass transfer rate based on the thermal time-scale (Eq.~(\ref{eq:tauth})) might be too simplified.
Therefore the upper limit of the donor mass obtained here also might be a tentative value. 
In order to investigate the precise RLOF rate, the response of the surface layer of $2 M_{\odot}$ star to mass loss is required. 
In general, further detailed treatments require additional parameters such as a pressure scale height \citep[for example]{R88}, and is out of the scope of our present study.
According to detailed computations \citep{P02}, however, the mass transfer via RLOF proceeds within nearly thermal time-scale for compact binaries with donor stars $\lesssim 4 M_{\odot}$.

The X-ray emission mechanism from the accreting neutron star with nearly the Eddington rate is old problem but complex and still poorly understood \citep{BS75, E14, M15}. 
The newly found ultra-luminous X-ray source powered by accreting neutron star in M82 \citep{B14} made such a problem confusing. 
A super-Eddington accretion onto a strongly magnetic neutron star could be closely connected with jet formation and/or strong beaming \citep{DPS15, D16, M15}. 
In this topic, further observational and theoretical improvements are required. 
In the next section, we discuss the recent study about the propeller effect in the ultra-luminous accreting neutron stars given by \citet{D16}.

\section{Discussion}

One of the prominent differences between LMXB and HMXB is a magnetic field of its accreting neutron star. 
That is, in general, a neutron star in LMXB has much weaker magnetic field strength (typically, $B \sim 10^{9} \rm{G}$), while a neutron star in HMXB has a magnetic field as strong as $\sim 10^{12} \rm{G}$ \citep{B97, H14, K14,BK16}. 
Hence, if we could know the magnetic field strength of the neutron star in M31 XBP, we could also infer the category of this system. 
In \citet{E15}, they assume a rotational equilibrium and estimate the neutron star magnetic field as $B \approx 10^{12} \rm{G}$ in this system. 
This magnetic field strength is much stronger than typical LMXBs which harbor old neutron stars; it is much similar to the young neutron stars in HMXBs. 

\begin{figure}
 \begin{center}
  \includegraphics[width=8cm]{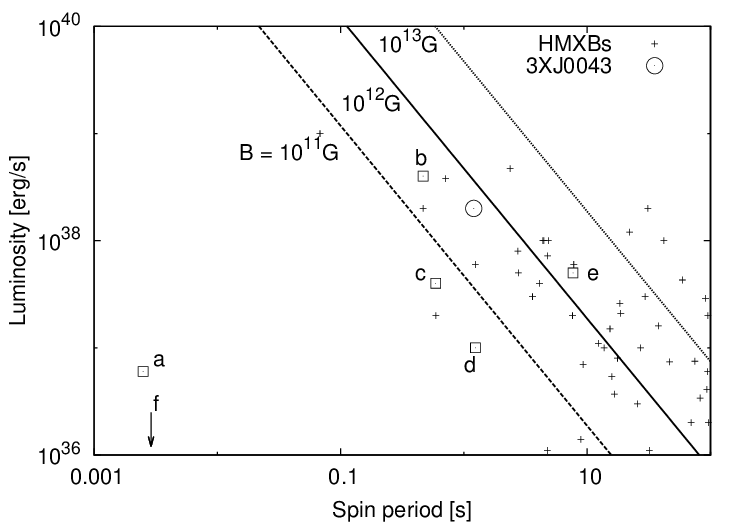} 
 \end{center}
\caption{
The limiting lines of propeller effect for different neutron star magnetic fields. 
The position of M31 XBP is shown by an open circle. 
Selected samples of HMXB (crosses) and LMXB/IMXB (squares) are also shown.
LMXB/IMXB samples denote (a) SAX J1808.4-3658 \citep{CM98} , (b) GRO J17442-28 \citep{L96,F96} , 
(c) 4U 1822-37 \citep{JK01}, (d) Her X-1\citep{T72, B97}, (e) 4U 1626-67 \citep{C98,O98}, 
(f) PSRJ 0636+5129 ($L_{\rm{X}} \sim 10^{31 \pm 2} \rm{erg \, s}^{-1}$, \cite{S16}), respectively.
}\label{fig:3}
\end{figure}

In the XMM Newton observations, small changes of spin period between three epochs have been reported. 
That is, the spin period of the neutron star in M31 XBP was $1.203892 \pm 0.000001$s in the first epoch, $1.203644 \pm 0.000003$s in the second epoch, and $1.2037007 \pm 0.0000003$s in the third epoch \citep{E15}. 
Though they are small, these changes of $P_{\rm{spin}}$ could be considered as a sign of the spin-up/down episodes.  
If so, it means that this system could be just near from the limit of propeller effect \citep{IS75}. 
In Fig.~\ref{fig:3}, we show the limiting lines of propeller effect which can be obtained from the following condition: 
\begin{equation} 
P_{\rm{s,crit}} = 81.5 \left( \frac{B}{10^{12} \rm{G} } \right)^{16/21} 
\left( \frac{L_{\rm{X}}}{10^{36} \rm{erg \, s}^{-1}} \right)^{-5/7} \rm{s}, 
\label{eq:prop} 
\end{equation} 
where $B$ is the dipole magnetic field, and $L_{\rm{X}}$ is the X-ray luminosity
\citep{I03,RP05}. 
Since this system is emitting X-rays with almost the Eddington luminosity, the propeller effect had been switched off at the time of observations, but not so far from the propeller limit. 
From this fact, the magnetic field strength of the neutron star in this system can be roughly estimated as $B \approx 7 \times 10^{11} \rm{G}$ (see Fig.~\ref{fig:3})
\footnote{
This value is a little bit smaller than the magnetic field strength given by \citet{E15}; $B \approx 1.3 \times 10^{12} \rm{G}$. 
In \citet{E15}, they assumed that the neutron star is in spin equilibrium. 
On the other hand, Eq.~(\ref{eq:prop}) comes from a different condition that the supplied rotational kinetic energy into the magnetosphere is larger than the radiated energy loss \citep{I01, DP81}. 
}. 
This magnetic field is rather similar to HMXBs. 
It means that the neutron star in M31 XBP is relatively young; it supports the prediction that this system is an intermediate mass X-ray binary.
In Fig.~\ref{fig:3}, we also show the position of some LMXBs.
In general, however, since neutron stars in LMXBs do not show pulsations (perhaps because of too weak magnetic field), spin periods of neutron stars in most of LMXBs are unknown. 
Hence, shown systems in this figure may be rather peculiar cases as LMXBs.

On the other hand, \citet{Z16} suggested that this system is a member of LMXB and proposed another scenario to avoid the confliction between strong magnetic field and age of LMXB.
They argued that the neutron star in this system was born via an accretion induced collapse recently; hence it could still hold a strong field.
Also in this scenario, however, the lower mass limit of $\sim 1.5 M_{\odot}$ for the donor could not be avoided, since this is dictated by the evolution time required to leave the main sequence and expand up to Roche radius.

The spin of this neutron star is slower than the propeller limit given in Eq.~(\ref{eq:prop}).
This means that this neutron star is spinning up due to the angular momentum accretion. 
Hence, continuous observations would give important information about the variation of the spin. 
The observed data of $\dot{P}_{\rm{s}}$ would give us further reliable evaluations of stellar magnetic field.
Since the magnetic field strength is associated with the system age, from these data we would be able to disclose the identity of this system. 
In \citet{D16}, a large fluctuation of the X-ray luminosity in accreting magnetic neutron star is discussed in detail, taking the super-luminous neutron star M82 X-2 as an example. 
They argue the importance to consider balances between (i) the magnetic radius of the neutron star and the corotation radius, 
(ii) the gas-pressure dominant region and the radiation dominant region in the disk, to understand the X-ray luminosity changes for super-Eddington sources. 
At present, M31 XBP is not an ultra-luminous source, and its magnetic field is possibly smaller than the quantum limit. 
However, if M31 XBP shows the same kind of large fluctuations of X-ray luminosity in a long term observations, further discussions about the neutron star magnetic field would be possible.

The donor mass of X-ray binary is important information to understand massive star formation mechanisms. 
An X-ray binary is one of evolutionary phases of binary systems containing at least one massive component which evolves to a neutron star after a supernova explosion. 
Metal ejection due to massive stellar wind and supernovae play critically important roles in galactic chemical evolutions.
However, properties of massive binaries including the formation rate, binary fractions, and mass-ratio are still unknown \citep{SE11}.

One of the reasons of these uncertainties is that massive stars have only short life-time, and hence observational chances are limited. 
Additionally, the population of massive stars itself would be very small: assuming a power low initial mass function, the population of massive stars decreases as $M^{-2.3}$ \citep{S55,K01}. 
Then, the population of X-ray binaries could play a role to complement our understanding of such massive star binary populations.
In general, it is suggested that a massive star tends to construct a binary system \citep{B98, S12}. 
The population of IMXB/HMXB would afford a collateral evidence of such a tendency. 
Also such a ratio would be a key to obtain the unknown mass-ratio of massive binary systems.
In fact, the observed number of IMXB is much less than that of HMXB. 
Since, however, intermediate mass stars are not suffered from mass loss due to stellar wind, a neutron star component hardly accretes and emits X-ray radiations. 
In addition to this, even if a system evolve into an X-ray emitting regime due to donor expansion, its life-time would be very short \citep{P02}. 
Hence, it is difficult to know the actual population of IMXBs. 
In order to obtain the actual number ratio between HMXB and IMXB, our stock is too small, and we need much more samples. 
Earning enough observational data set that can be used in statistical analysis, by comparing with the results of theoretical works including population synthesis, we would be able to improve our understanding about massive stellar populations and binary fractions.

\section{Conclusion}

In this study, we have examined the binary component of the newly found accreting X-ray binary in M31, by using a stellar evolution track.
Since there is a strong condition that the donor star should fill its Roche lobe before the cosmic age, the minimum mass of the donor star is $1.5 M_{\odot}$. 
Though the upper limit on the donor mass could be around $2 M_{\odot}$, although its exact determination is subject to residual uncertainties.
It strongly suggests that this system is an intermediate mass X-ray binary as like a prototypical case of Her X-1. 
This idea is supported by relatively strong magnetic field of the neutron star deduced by assuming spin equilibrium {\it and} that the source is close to the transition to propeller. 
The magnetic field of the neutron star in this system is actually much stronger than the typical LMXB neutron stars, and rather similar to young neutron stars. 
In future, we would be able to obtain much more concrete figure of this system, through further observations, such as spin evolutions.





\end{document}